# FAULT DIAGNOSIS USING CLUSTERING. WHAT STATISTICAL TEST TO USE FOR HYPOTHESIS TESTING?

Nagdev Amruthnath[1]and Tarun Gupta[2]

[1]Department of Industrial Engineering, Western Michigan University, Kalamazoo, MI, 49008, USA
[2]Department of Industrial and Entrepreneurial Engineering, Western Michigan University, Kalamazoo, MI, 49008, USA

## ABSTRACT

*Predictive maintenance and condition-based monitoring systems have seen significant prominence in recent years to minimize the impact of machine downtime on production and its costs. Predictive maintenance involves using concepts of data mining, statistics, and machine learning to build models that are capable of performing early fault detection, diagnosing the faults and predicting the time to failure. Fault diagnosis has been one of the core areas where the actual failure mode of the machine is identified. In fluctuating environments such as manufacturing, clustering techniques have proved to be more reliable compared to supervised learning methods. One of the fundamental challenges of clustering is developing a test hypothesis and choosing an appropriate statistical test for hypothesis testing. Most statistical analyses use some underlying assumptions of the data which most real-world data is incapable of satisfying those assumptions. This paper is dedicated to overcoming the following challenge by developing a test hypothesis for fault diagnosis application using clustering technique and performing PERMANOVA test for hypothesis testing.*

## 1. INTRODUCTION

Machine maintenance has been one of the critical components of the manufacturing industry. It is estimated that US industry spends about $200 billion a year in machine maintenance [1]. Over the years, various maintenance techniques have been proposed such as preventive maintenance (PM), reliability-centered maintenance (RCM), and predictive maintenance (PdM) [2]. Every new method has proved to be much more efficient and providing a competitive advantage in both quality and cost of the product. Most manufacturing industries today use preventive maintenance due to its high maintenance efficiency. In the last couple of years, predictive maintenance has seen a significant acceptance in manufacturing due to accessibility and handling the manufacturing process data in real time, inexpensive sensors [3]and software that is capable of handling big data and performing real-time data analytics. Today, predictive maintenance involves collecting machine data, performing signal processing, early fault detection, fault diagnosis, time to failure prediction, maintenance resource optimization and scheduling by using concepts from statistics, machine learning, and data mining [4]. Based on the following components, various architectures and processing models have been proposed over the years, and most of the architectures use supervised learning[5][6]. In a real-world manufacturing environment, most machines work in fluctuating environments where, the operating temperature

varies based on the type of products, fluctuating mechanical vibrations, fluctuating sound, and other industrial noises. This can contribute to adding noise to the data. In these cases, a supervised learning model's accuracy deteriorates as it is trained for a specific environment with specific failure modes. To overcome this challenge, supervised learning techniques can be trained with data which considers all the identified factors. This approach can sometimes be unrealistic to achieve, and in some cases, it can be expensive[7]. Also, if the model is not trained with a new type of fault, then the machine would not be capable of detecting these faults. An alternative approach to overcome this challenge is to use unsupervised learning models. This technique can be used for both early fault detection and fault diagnosis. In recent times, this technique is used to detect the severity of the machine state. Some of the conventional clustering models used for these applications are k-means [4], Gaussian mixture modeling[4][8], self-organizing map [9], neural networks and random forest clustering [10].

Clustering analysis is a data mining technique where similar data are grouped based on the underlying information within those groups. This technique is generally used to provide deeper insight into the data. Clustering is extensively used in different applications such as customer segmentation in marketing [11][12], big data analytics[13], data mining[14], fraud detection[15], and sound analytics[16]. Some of the most commonly used cluster validation techniques are internal cluster validation and external cluster validation [17]. In internal cluster validation, the structure of the cluster, separation between and within clusters are determined based on different indices to validate the significant separation between groups[18]. In external cluster validation, the clustering results are verified with known class information to evaluate the effects[19]. Any model developed for fault detection, and diagnosis can be substantiated and validated using the class labels if available or by cross-referencing with the actual events such as maintenance records (also called an external cluster validation). This is one of the most common techniques that have been widely used. With this technique, different statistics such as accuracy, kappa value [20], and No information rate (NIR) is calculated to compare the reliability of the model[21]. In cases where no label information is available for fault diagnosis, then a hypothesis can be designed along with an appropriate statistical test if, there is any significant difference between the groups and conclude that these groups are significant. Today, some of the most commonly used techniques are an analysis of variance (ANOVA), multivariate analysis of variance (MANOVA), and t-test. These tests provide a statistical test on the means of the test groups and a post hoc test to compare which pairs are significantly different. These techniques require some assumption regarding the data such as [22]

- data follows a normal distribution
- linearity in data
- all groups have the same variance
- homogeneity of covariance
- the presence of outliers

The issue with real-world data is to satisfy these assumptions to fit the data to a model. In most cases, there is the presence of outliers in data or a valid data point resembles as an outlier, most data don't always follow a normal distribution, and non-linearity is commonly observed. Over the years different methods have been proposed to overcome these challenges such as outlier detection test to detect and remove outliers from the data[23], and transformation techniques to transform non-normal data to normality[24]. By removing outliers and transformation, there is a possibility of losing significance in the data. To overcome these problems, this paper proposes a new technique of using Permutational ANOVA proposed by Anderson for testing the statistical significance of fault diagnosis clustering models[25]. This statistical test is widely used in various applications such as ecology[26], biology, marine biology[27]and microbial study[28]. One of the main advantages of this technique is that there can be more factors than the number of samples and there is no effect on the presence of "zeros" in the data.

## 2. LITERATURE REVIEW

One of the most commonly used statistical tests for hypothesis testing is t-test when testing for two groups' means and when there are more than two groups, ANOVA is used. ANOVA was first proposed by Ronald Fisher, a biologist, and a statistician. ANOVA is widely used for comparing the means of two or more groups for statistical significance [22]. This technique is extensively used in testing experimental data. When the data is collected, a hypothesis can be designed where the null hypothesis is "there is no significant difference in the means between the groups." Hypothesis testing and its significance level limit the Type I error rate (false positives). But, Type II error (false negative usually depends on sample size, significance level, and effect size. In cases of multiple response variables, MANOVA is typically used.

Fault diagnosis and severity detection are one of the critical components of predictive maintenance. With proper diagnosis technique, the process of fault investigation can be eliminated during the maintenance schedule and hence the overall time to repair can be minimized. With the severity detection techniques, the maintenance schedule can be optimized to minimize the cost of maintenance. Today, diagnosis technique is performed using machine learning[4], deep learning, statistics, and data mining techniques using labeled data. To have a robust and reliable model that is capable of working across different environments and scalable to different applications, unsupervised clustering techniques have shown favorable results in the current literature [4]. Clustering is used in various applications such as customer segmentation[12], image classification[29], data mining[14], production flow analysis [30], and pattern recognition. One of the key challenges that exist with this technique is to properly validate the results using the appropriate statistical test to test if, there is a significant difference between the clusters. The other problem is, most statistical analyses that are commonly used today have a set of assumptions and most real-world data fail to satisfy those assumptions. One of the widely encountered assumptions is a test for normality where parametric tests assume that the data follow a normal distribution. When these fail, then the other option is to use a non-parametric test.

## 3. HYPOTHESIS TESTING

In this research, a condition based monitoring system was set up for an industrial furnace cooling fan. Accelerometers were mounted to monitor the degradation cycle of the machine on radial and axial positions [2]. The vibration data is collected using wireless vibration sensors mounted on x-axis and y-axis as shown in Figure 1 and Figure 2. The data sampling frequency was 2048 Hz [4]. The vibration data were collected every 10 minutes for five months. This data was processed continuously using machine learning techniques for feature extraction and fault diagnosis.

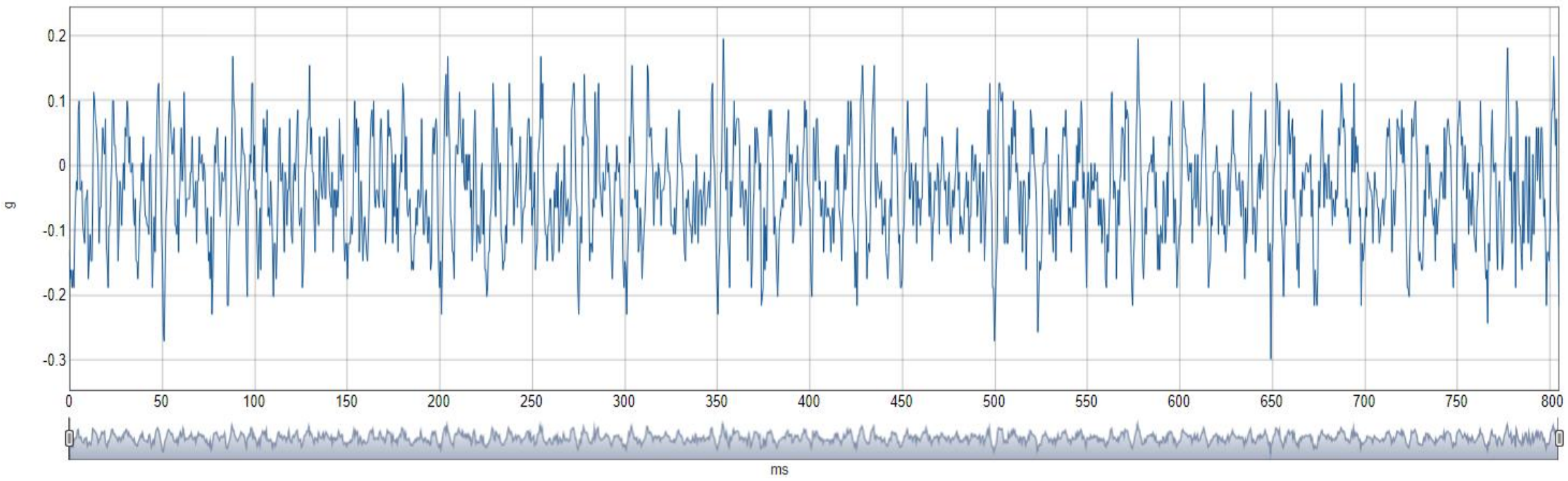

Figure 1: Raw vibration data sample for Y-axis in terms of g

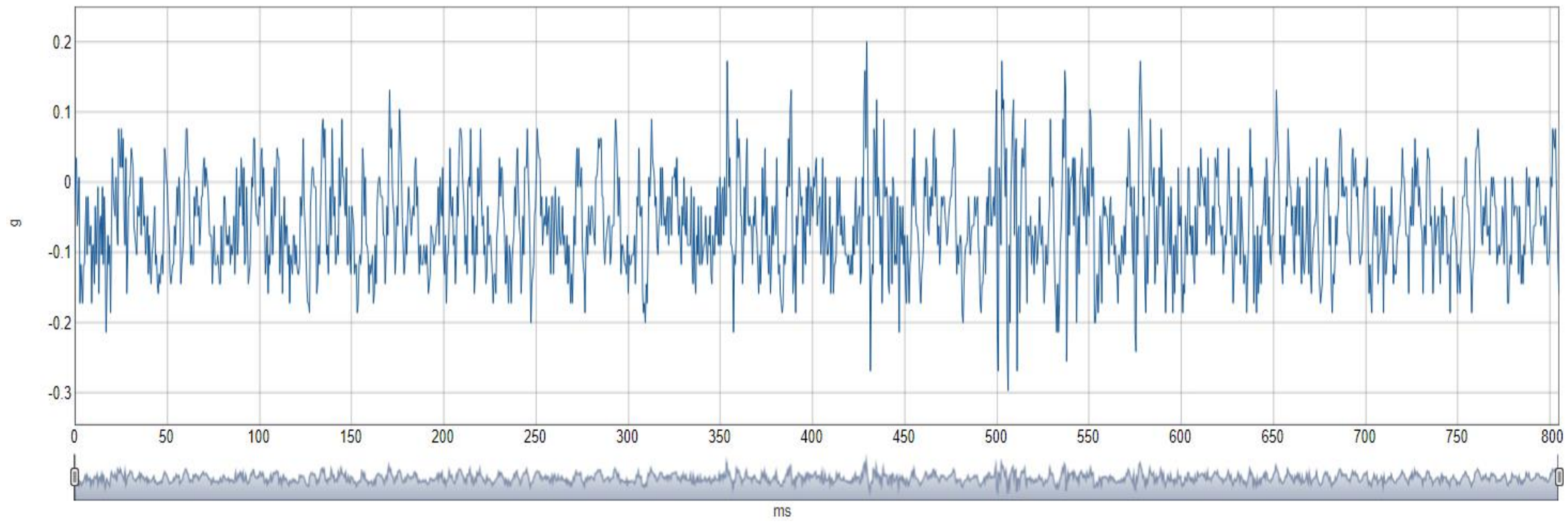

Figure 2: Raw vibration data sample for X-axisin terms of g

### 3.1. Feature Extraction and Analysis

Feature extraction in machine learning is a process of extracting significant attributes of the data. These characteristics can be statistical features, domain-specific features or both. In this research, the statistical features of the vibration data were extracted which include mean, median, min, max, kurtosis, skewness, standard deviation, RMS, and range [31]. These features were obtained for both x-axis and y-axis data. The collected vibration data was in time series, and this information was transformed into the frequency domain using fast Fourier transforms as shown in Figure 3 and Figure 5 [32]. The machine's operating frequency was identified to be 26.1Hz, and hence, significant features in the frequency domain were collected at its operating frequency.

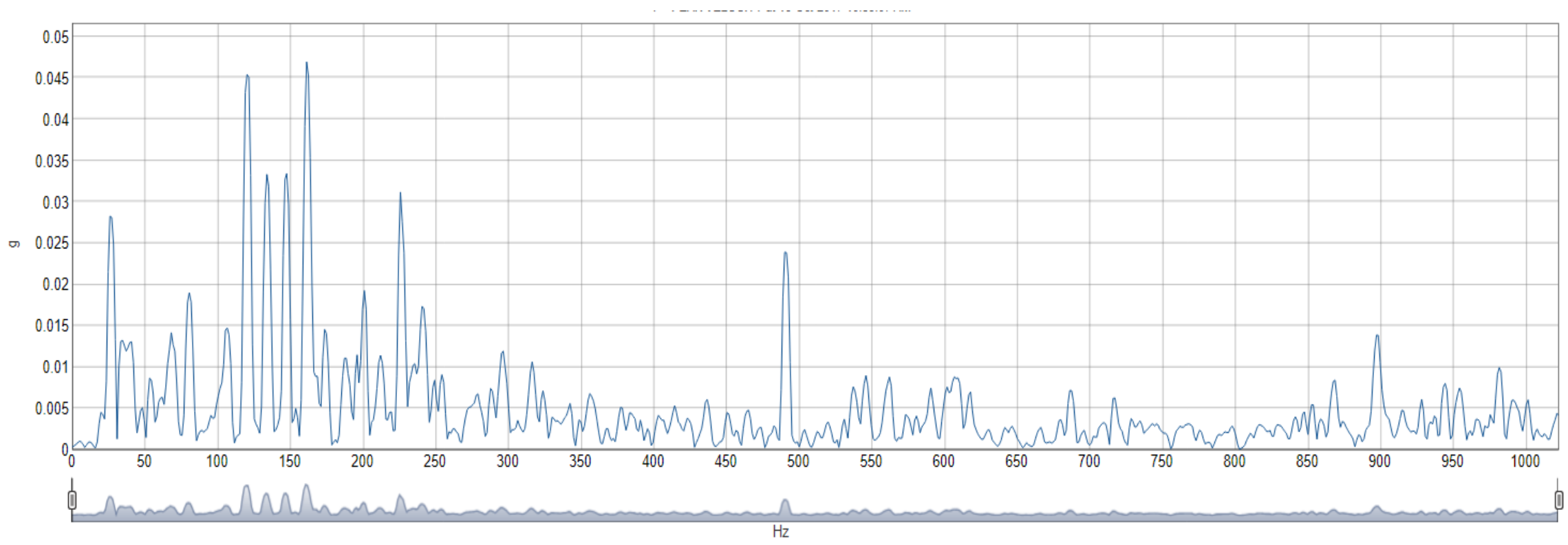

Figure 3: FFT result of vibration data for Y-axis

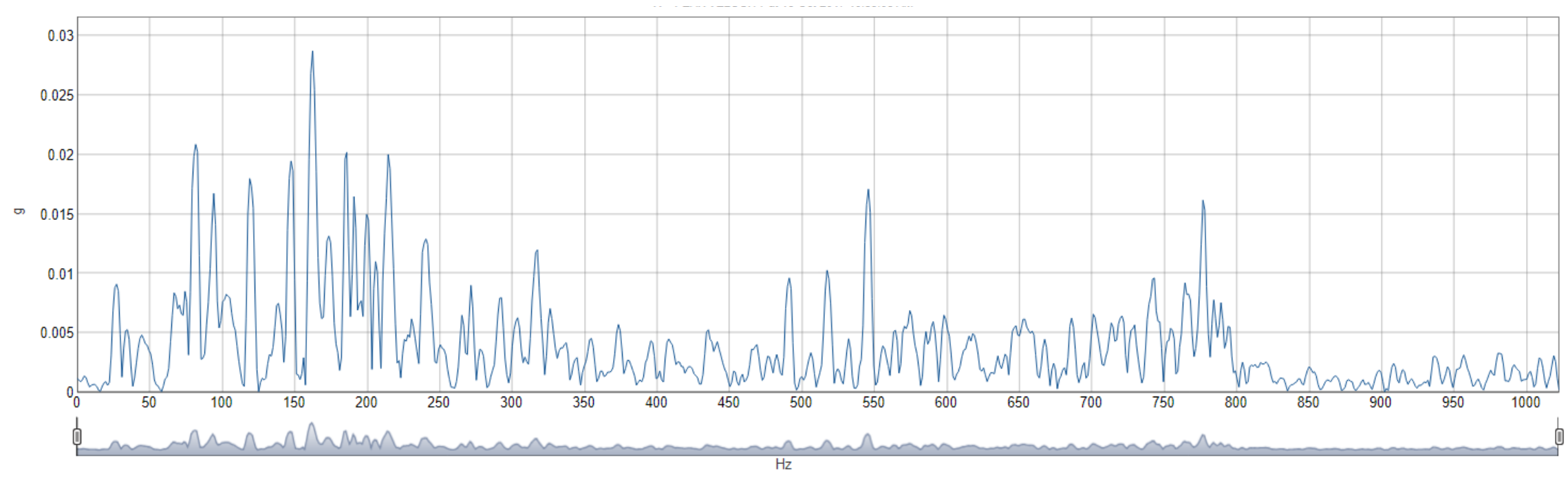

Figure 4: FFT result of vibration data for X-axis

Feature data analysis can be performed using various techniques that are available today. Some of the most commonly used methods are histograms and density plots to identify the distribution in the data, data uniqueness, and variance analysis to identify all the features collected for significant and unique, box-plots to determine any presence of outliers. In this research, upon performing density analysis for various features multi-model density plots were identified. This information will later be used in our clustering assumption. By carrying out the test for unique values, we were able to determine that all the features used in the data had unique values. By plotting the box-plots for normalized data, we were able to identify the range of feature data and the presence of outliers as shown in Figure 5.

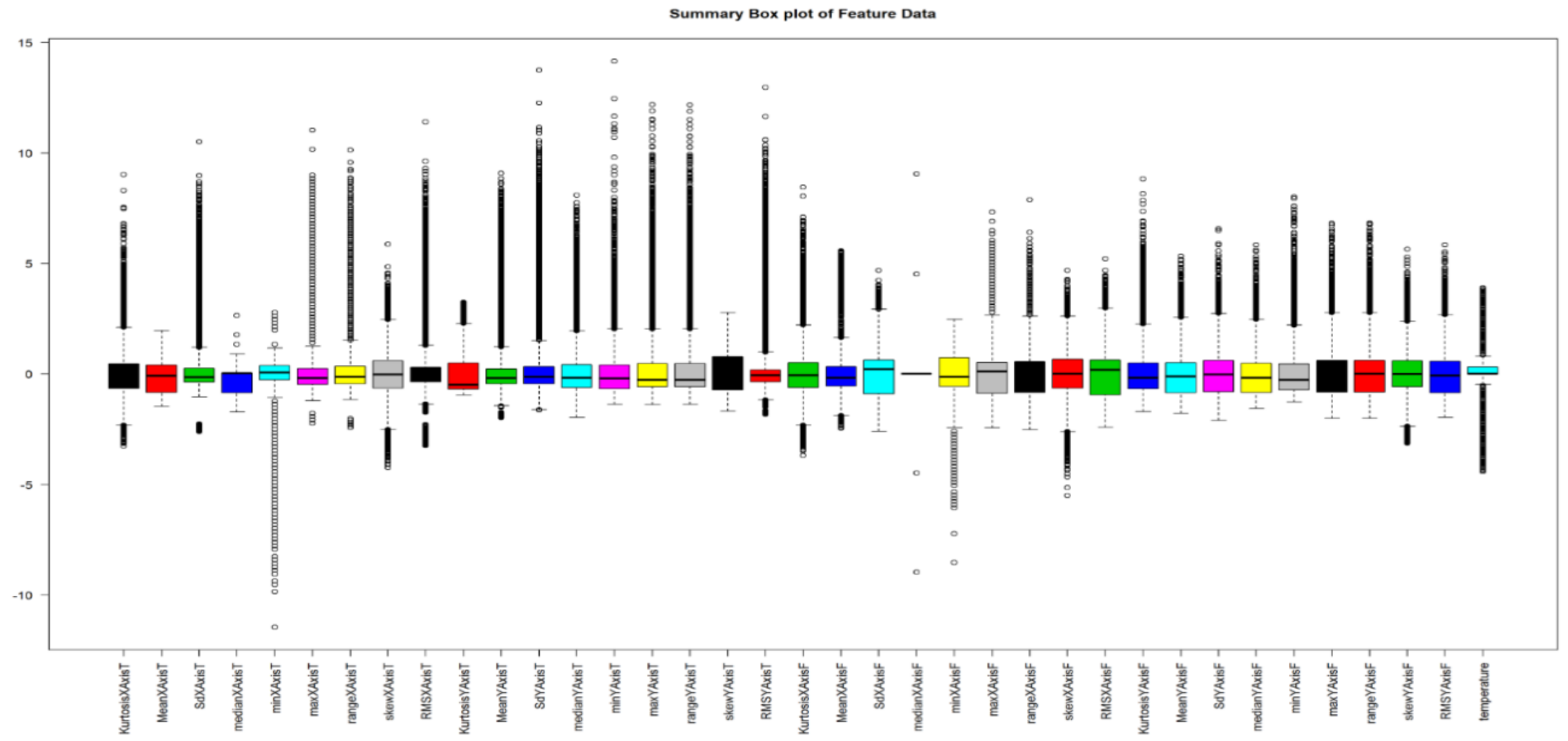

Figure 5: Box-Plot for normalized feature data

## 3.2. Cluster Analysis

Clustering analysis is an unsupervised learning technique that is used to group similar types of data. The data can be clustered based on density[33], distance[34], p-value or various other practices. Some of the most commonly used models in clustering for machine state detection are k-means [35], hierarchical clustering, Gaussian mixture model (GMM) based clustering [4], Bayesian mixture model-based clustering, density-based clustering and fuzzy clustering [32]. GMM clustering technique was used for clustering the data into distinct groups. GMM is a probabilistic model that assumes all the feature data are generated from a mixture of a finite number of Gaussian (normal) distributions with unknown parameters[8].

In clustering, there is no definite answer for identifying the number of clusters. In this case, the number of clusters was unknown. To determine the optimal number of clusters, different techniques such as silhouette width, AIC [36], BIC [37], and within the sum of square (WSS) [4] methods are available. In this research, the optimal number of clusters were identified using WSS technique as the dimensionality of the data increase, the performance of AIC and BIC techniques reduces [38]. The obtained result was confirmed by generating the average silhouette width for k-number of clusters and identifying the point where it provides the maximum separation. The results of the WSS technique is as shown in Figure 6,and average silhouette width for k-clusters is as shown in Figure 7. Here, we can observe that there is no significant difference in WSS after cluster six unlike the previous number of clusters. Using this information, we can confirm that the optimal number of clusters is six. Average silhouette width is calculated for k clusters as shown in Figure 8. The maximum separation is observed at k= 3 clusters. Upon more profound observation, we can also observe that there is no significant difference in average separation between k=2 to k=8. This information also confirms that the optimal number of cluster identified in the data is six. Upon determining the optimal number of clusters, the feature data was clustered using GMM clustering. The results are as shown in Figure 9.

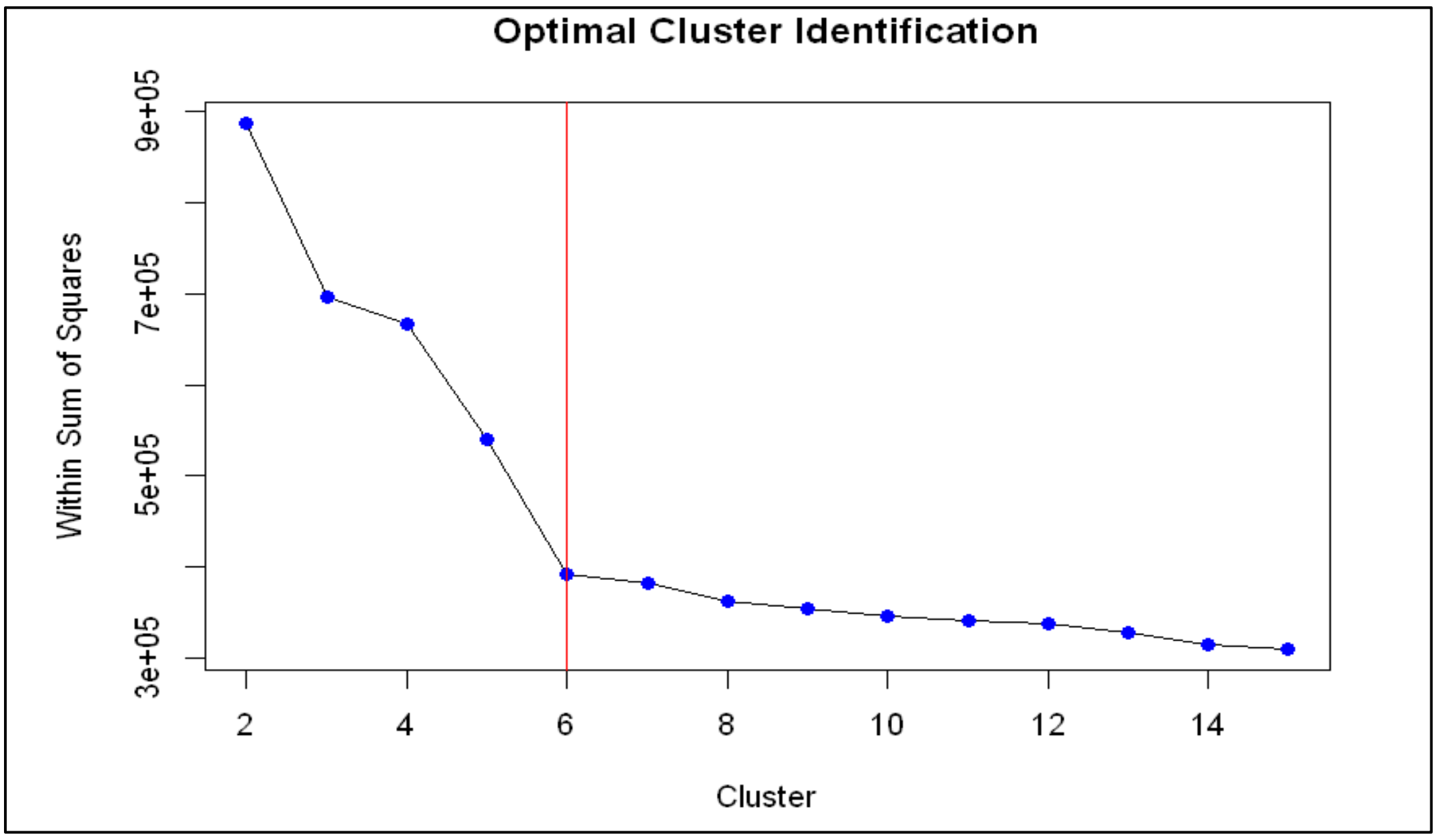

Figure 6: WSS and elbow technique for identifying the optimal number of clusters

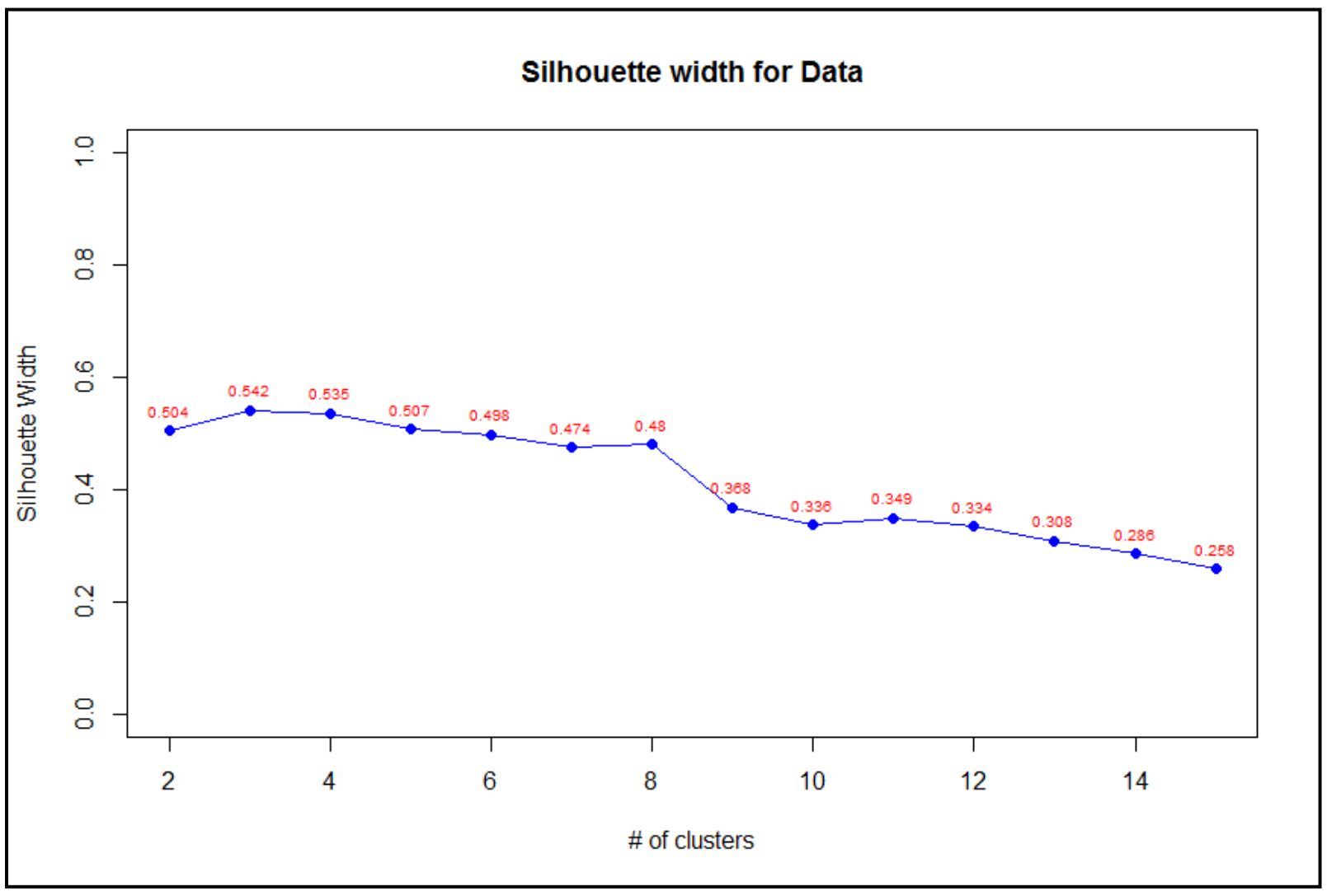

Figure 7: Average silhouette separation for 2 to 15 clusters

From the maintenance records, it was identified that cluster 1 was the normal operating state of the machine after fan replacement. Cluster 2 was the shaft displacement failure mode. Cluster 3 was identified to be furnace power off state. Cluster 4 was the normal state of the machine. Cluster 5 was the repaired state. Cluster 6 was identified to be imbalance state.

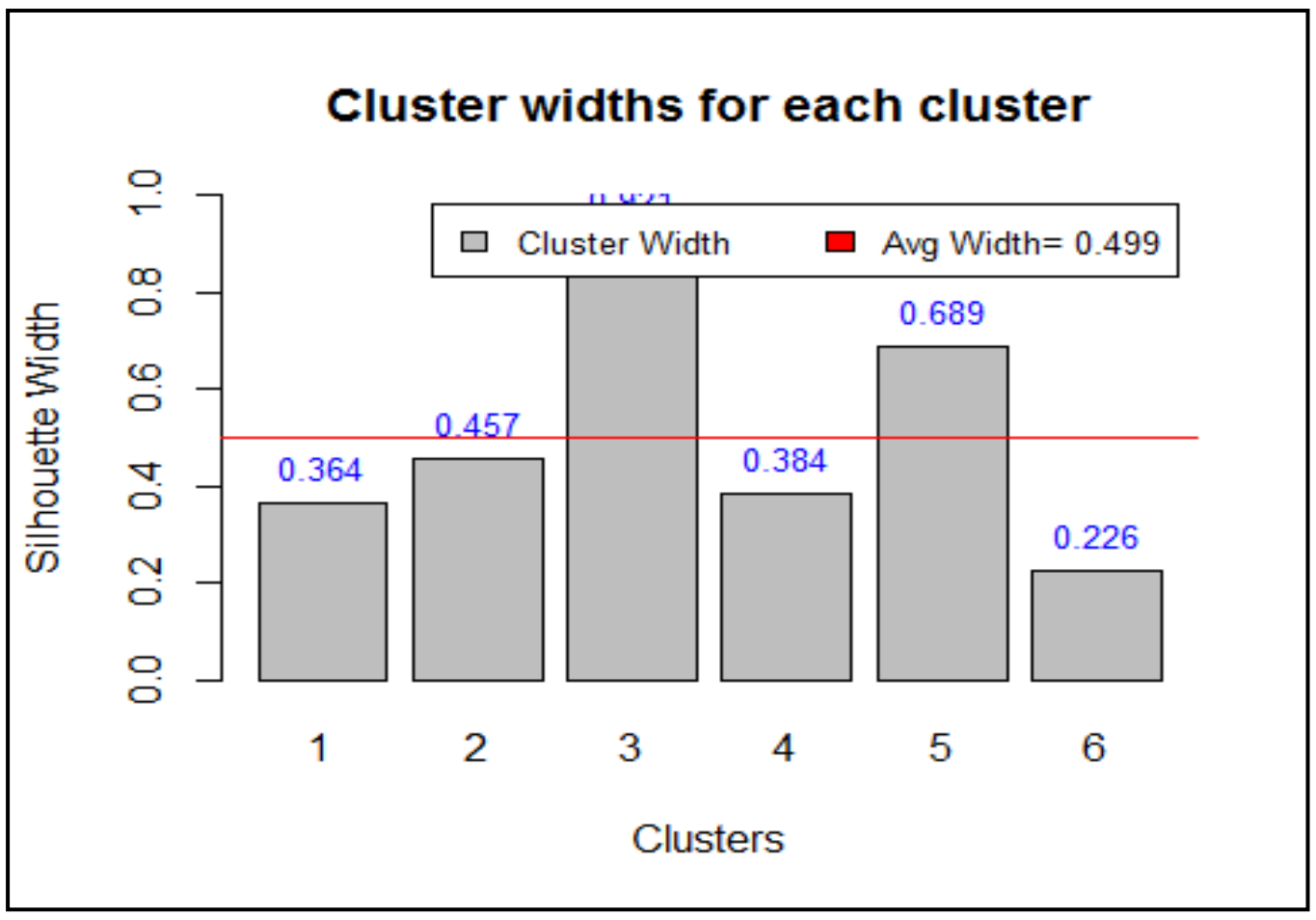

Figure 8: Cluster widths within clusters

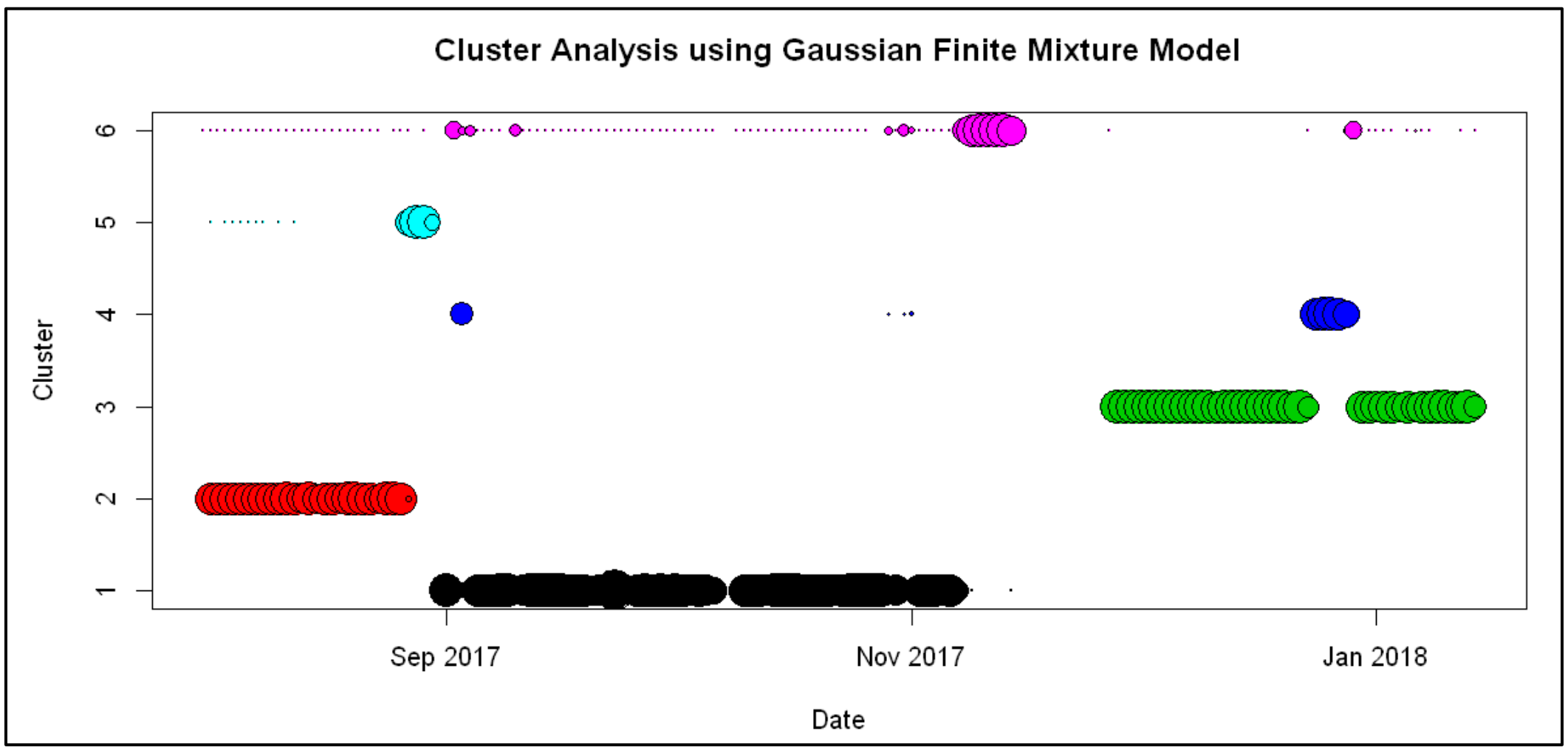

Figure 9: GMM cluster results in time series

In the cluster data, we can observe that there was a presence of an imbalance in the assembly since the beginning of data collection. On August 26, an abnormal state of the machine was identified while increasing the severity of the abnormality. On September 1, upon scheduled maintenance, it was identified that the shaft connecting the gear assembly and the fan were broken. Upon replacement of the shaft assembly, a new state was formed representing the repaired state. On November 14, we can observe the severity of the imbalance increase. The entire assembly was replaced during the scheduled maintenance. Upon replacement of the entire assembly, the imbalance state did not occur and formed a new cluster for the new assembly. The machine was shut down from December 24 to December 30. This was accurately recognized by the clustering model by forming a new cluster. A brief imbalance cluster formation is noticed before shutdown and after turning on the machine briefly. This imbalance condition was confirmed to be a normal operation for the machine.

### 3.3. Visualization of Data in Low Dimensional Space

In this research, to study the groups within the data a process of dimensionality reduction was required. This was achieved using principal component analysis (PCA)[39]. Using PCA the variance within the data was maximized, and the first three components were chosen based on scree plot as shown in Figure 10 to visualize the data in 3-dimensional space to detect any groups as shown in Figure 11 and visualize the clustering results in 3-D space as shown in Figure 12. In Figure 11, we can observe that there is a presence of groups and are identified using red circles.

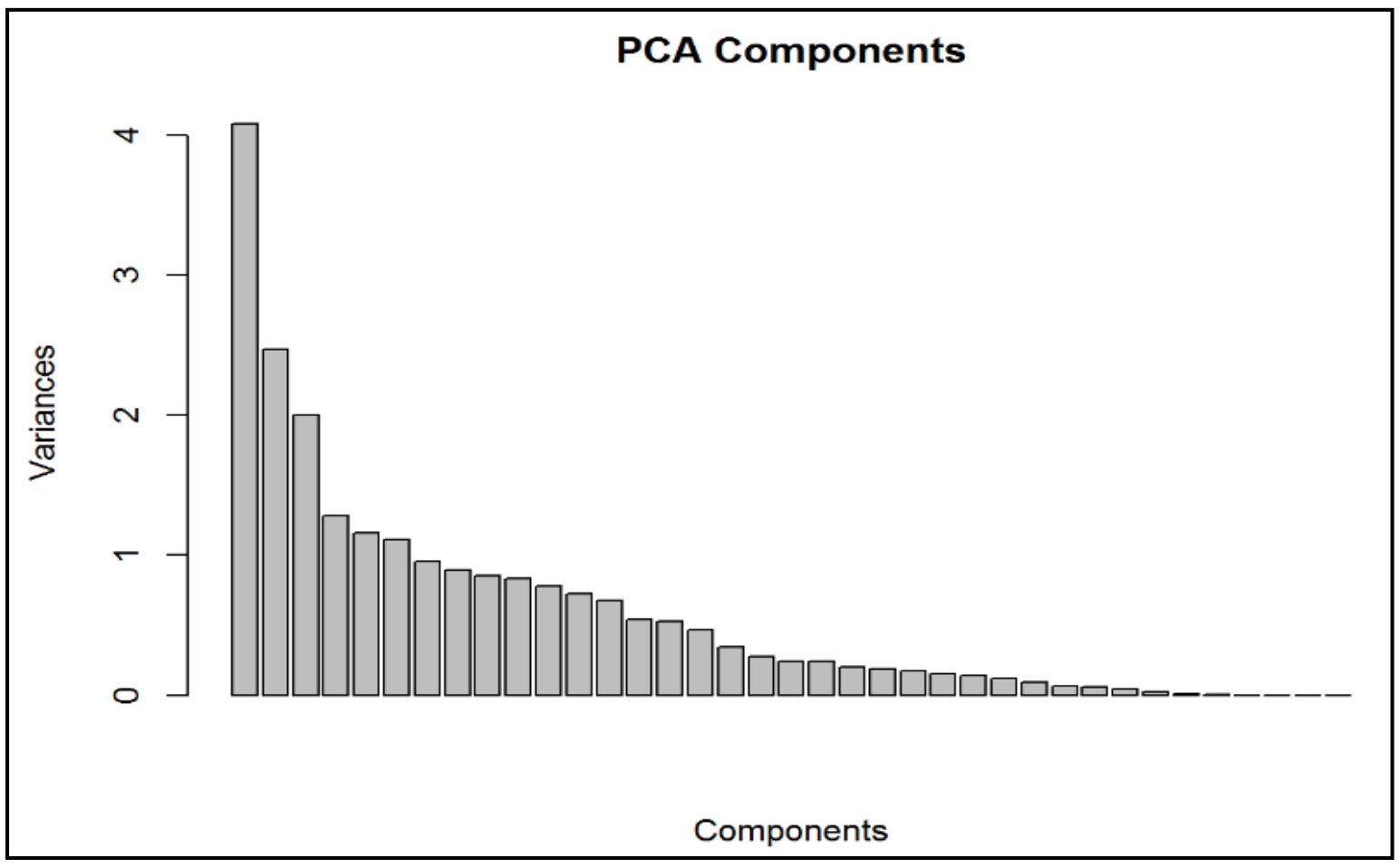

Figure 10: Variances for all the principal component vectors

Upon adding the cluster results to the 3D plot, we can observe all six groups in a lower dimensional space as shown in Figure 12. With the following visualization, we can subjectively validate the clustering results.

## 3.4. DESIGNING HYPOTHESIS

One of the essential tasks in this study was to group the data into its distinct groups; first by identifying the optimal number of clusters and then grouping using GMM technique. It was also very important that the clustered groups represent the actual state of the machine. The clustering results can be validated using the maintenance records and clustered groups in time series. It was also essential to show the significant difference between groups or clusters and this can be achieved by developing a hypothesis and performing the appropriate statistical analysis. In this research, a hypothesis was designed to identify the significant difference between groups.

***Hypothesis***

$H_0$: There is no significant difference in variation between the cluster groups
$H_a$: There is a significant difference in variation between the cluster groups

The hypothesis can be tested using various statistical tests such as ANOVA, chi-square test, and MANOVA analysis. Since the data in this research was a multivariate; multivariate data analysis was identified.

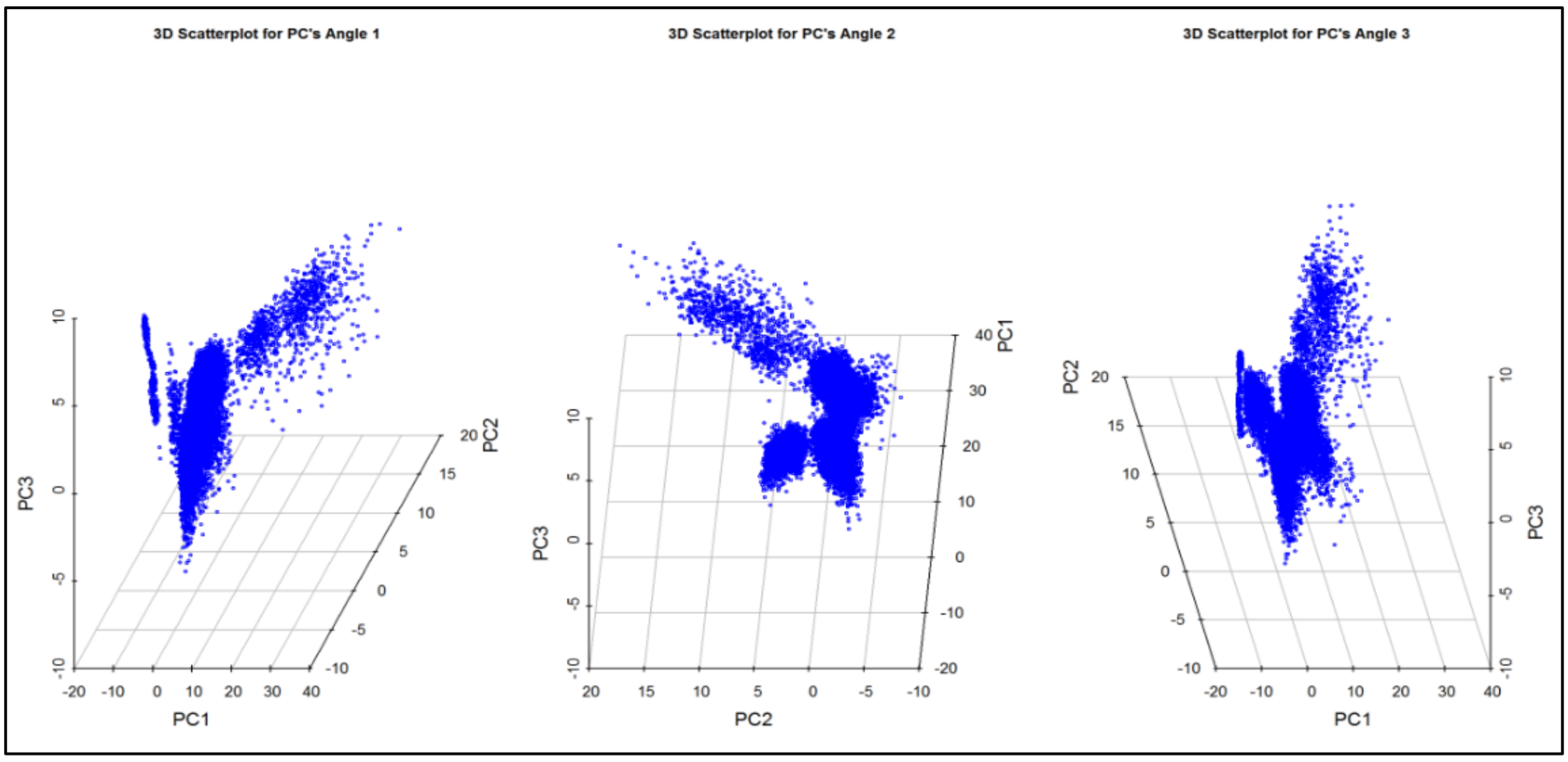

Figure 11: 3-D plot for the firstthree principal components at different angles

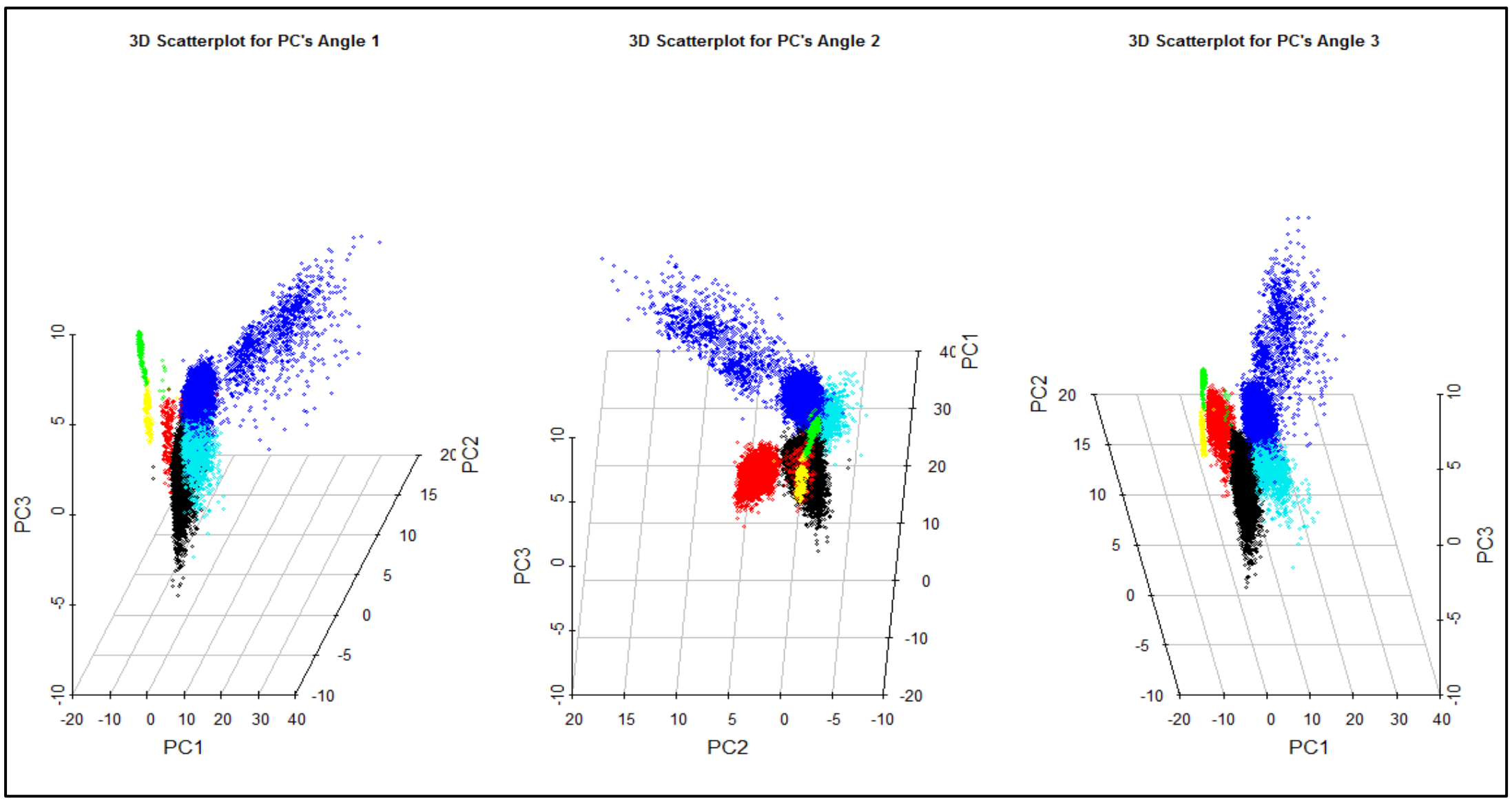

Figure 12: 3-D plot for the firstthree principal components at different angles with clustering results

# 4. RESULTS AND ANALYSIS

## 4.1. ONE-HOT ENCODING

One-hot encoding is a feature transformation technique where categorical variables are transformed indicating the state of the machine [40]. When this transformation is performed, univariate data is converted to multivariate response data. In classification problems, categorical variables are usually converted to a numeric variable so, the machine can understand. This also creates a bias during the training process where a categorical variable assigned with a higher number might be given greater prominence compared to a variable with a lowernumber where all variable should have equal prominence. In instances such as this case, one-hot encoding is performed to overcome that bias.

## 4.2. HYPOTHESIS TESTING USING MANOVA

### 4.2.1 Sampling

In statistical hypothesis testing it is important to sample data from the population. In this study, samples from each cluster (machine state) is extracted through randomization. The sample sizes for each cluster is 30.

### 4.2.2 Test for assumptions

MANOVA is a multivariate version of ANOVA analysis. Similar to ANOVA, MANOVA has a set of assumptions, and they are as follow

- Data follows a normal distribution
- Linearity in data
- All groups have the same variance
- Homogeneity of covariance

Along with the following assumptions, MANOVA is also very sensitive to outliers present in the data. If any of the assumptions fail to meet the criteria during the analysis, then the test would be an insignificant test. In this research, based on the density plots from the feature data, it was identified that the data follows multi-model distribution and this unmistakably fails to meet the assumption. Shapiro-Wilk normality test was used to test the normality of the feature data [23]. The null hypothesis and alternate for the test was as follows

*Hypothesis:*

$H_0$: the distribution of the data is not significantly different from the normal distribution
$H_a$: the distribution of the data is significantly different from the normal distribution

From a normality test, we observe that the p-value was less than 0.05 indicating that the distribution of the data is significantly different from a normal distribution and hence we reject the null hypothesis. From the results, it was determined that the data does not follow a normal distribution as shown in Table 1 and Figure 4.Also, in Figure 4, we can observe that there is a presence of outliers in the data.

Table 1: Shapiro - Wilk normality test

| Shapiro - Wilk normality test | |
|---|---|
| W | 0.96095 |
| p-value | 6.573e-05 |

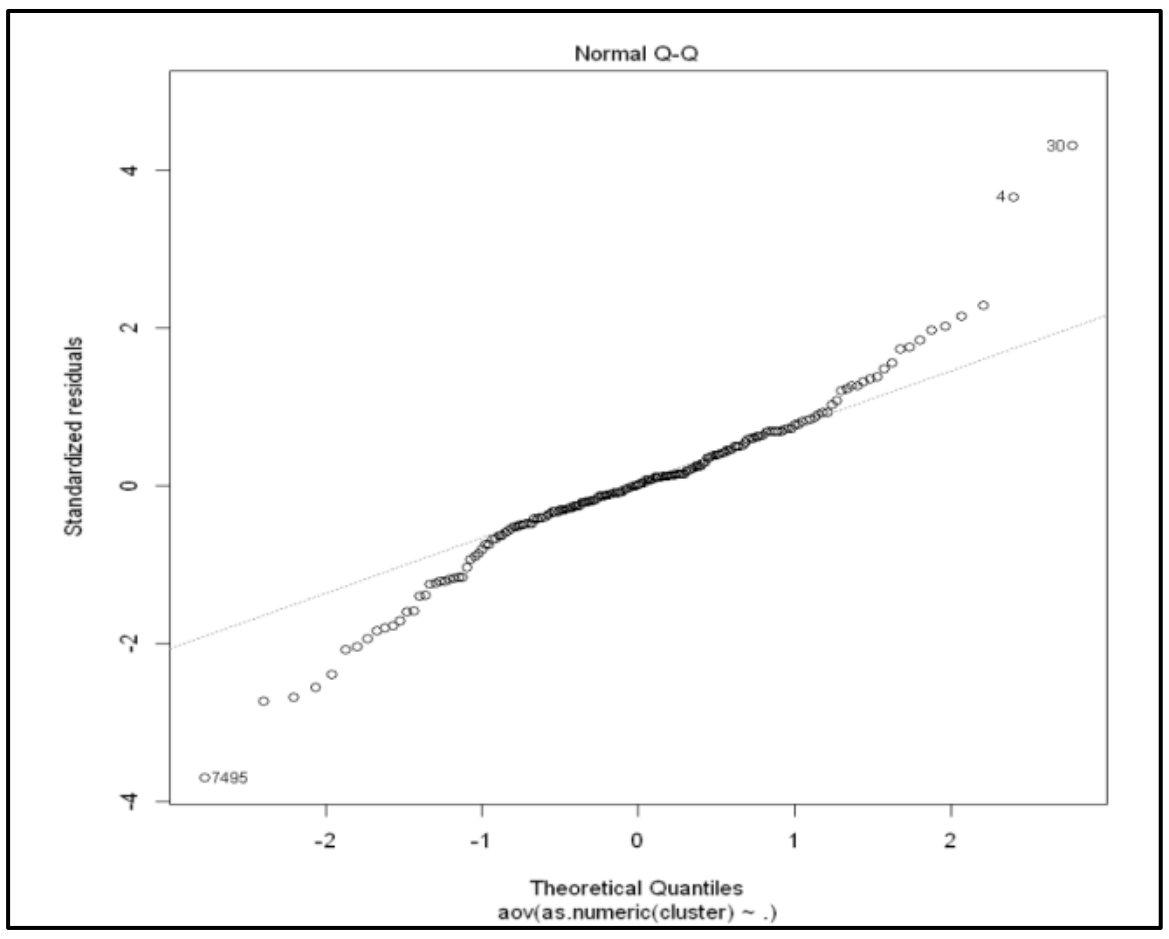

Figure 13: Test for normality in the feature data

Bartlett test of homogeneity of variances was used to test for equality of variances across groups against [41]. The null hypothesis for this test is as follows

*Hypothesis:*

$H_0$: the variance of two or more samples drawn from the same population have equal variance
$H_a$: the variance of two or more samples drawn from the same population do not have equal variance

Table 2: Bartlett 's test

| Bartlett test | |
|---|---|
| Bartlett 's K-squared | 93.919 |
| df | 5 |
| p-value | < 2.2e-16 |

From Bartlett 's test, we observe that the p-value was less than 0.05 indicating the variance of two or more samples drawn from the same population do not have equal and hence we reject the null hypothesis. From the results, it was determined that thereareat least two groups that have unequal variances.

Failing to meet the above assumptions are not unique to data obtained from real-world data. Domains such as ecology, microbial analysis, biology and many as such, it is often hard to satisfy the following assumptions. To overcome the above challenge, a new statistical analysis technique called PERMANOVA was proposed by Anderson in 2001[25].

## 4.3. Hypothesis testing using PERMANOVA

PERMANOVA also called per mutational MANOVA was proposed by Anderson to overcome the challenges of MANOVA [25]. The problems include the assumptions of MANOVA and sample size to be greater than the number of factors. This technique performs the analysis and partitions the sum of squares using dissimilarities. This technique also is formerly known as non-parametric MANOVA. Here, the inputs are linear predictors and a response matrix of an arbitrary number of columns; they are a robust alternative to both parametric MANOVA and to ordination methods for describing how variation is attributed to different experimental treatments or uncontrolled covariates [42].

Hypothesis:

$H_0$: the centroids of the groups, as defined in the space of the chosen resemblance measure, are equivalent for all groups
$H_a$: the centroids of the groups, as defined in the space of the chosen resemblance measure, are not equivalentto all groups

The results for hypothesis using PERMANOVA test is as shown in table 3.

Table 3: PERMANOVA statistical test results

| | DF | Sum Sq | F | Pr (>F) |
|---|---|---|---|---|
| Group | 26 | 132.964 | 45.92 | 0.001 |
| Residual | 153 | 17.03 | | |

In results from table 3, the permutation test was performed sequentially. Based on the results, we can identify that the p-value is less than the significant level of 0.05 indicating that there is a significant difference between groups. From the above test, we can conclude that the centroids of the groups, as defined in the space of the chosen resemblance measure, are not equivalent to all groups and reject the null hypothesis.

To observe the multivariate dispersion among the groups, a beta diversity test is performed. Beta diversity test is a multivariate analogof Levene's test for homogeneity of variances. Non-Euclidean distances between objects and group centroids are handled by reducing the original distances to principal coordinates. This procedure has latterly been used as a means of assessing beta diversity[42]. The hypothesis is as follows

Hypothesis

$H_0$: no difference between groups dispersion (here between machine states)
$H_a$: There is a significant difference between group dispersion

The results for testing the hypothesis are as shown in Table 4.

Table 4: The analysis of multivariate homogeneity of group dispersions (variances)

| | DF | Sum Sq | Mean Sq | F | N. Perm | Pr (>F) |
|---|---|---|---|---|---|---|
| Group | 5 | 100.45 | 20.099 | 19.87 | 999 | 1.20e-15 |
| Residual | 174 | 175.92 | 1.01 | | | |

Table 5: Pairwise comparison of permuted and observed P-values

| | | Permuted P-Values | | | | | |
|---|---|---|---|---|---|---|---|
| | | 1 | 2 | 3 | 4 | 5 | 6 |
| Observed P-values | 1 | | 0.91500 | 0.10600 | 1.0000e-03 | 5.0000e-03 | 0.001 |
| | 2 | 0.91744 | | 0.02600 | 1.0000e-03 | 1.0000e-03 | 0.001 |
| | 3 | 0.10402 | 0.02325 | | 7.0000e-03 | 1.0000e-03 | 0.001 |
| | 4 | 1.5085e-04 | 1.4035e-07 | 7.2922e-03 | | 1.0000e-03 | 0.001 |
| | 5 | 3.6307e-03 | 5.3810e-04 | 3.2128e-06 | 7.1421e-11 | | 0.138 |
| | 6 | 2.8593e-04 | 7.0673e-05 | 1.2442e-06 | 8.9320e-10 | 0.1359 | |

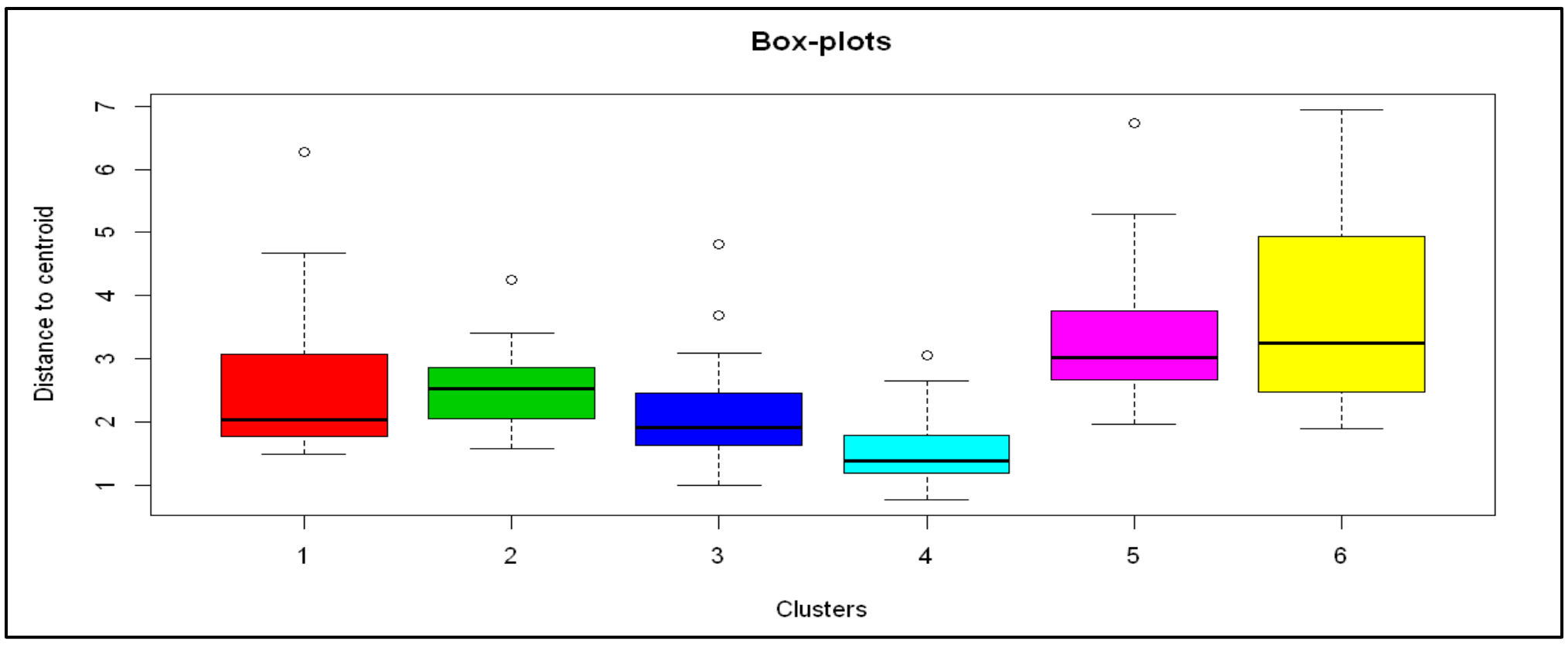

Figure 14: Box plot for clusters and distance to the centroid

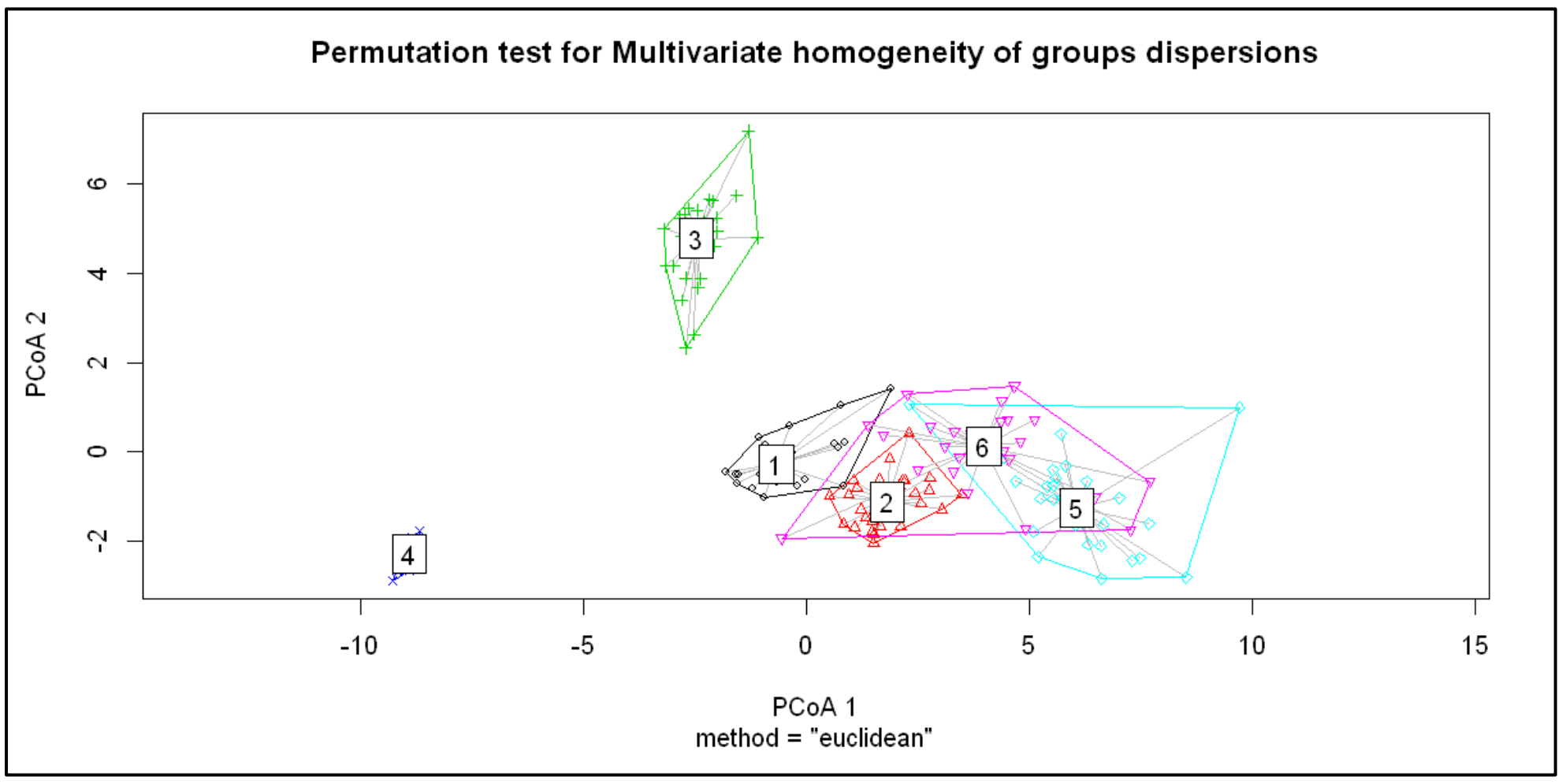

Figure 15: PCoA for clusters

In the results from Table 4, based on the p-value results we can observe that there is a significant difference between the groups and reject the null hypothesis. Since the test results are significant, a pairwise comparison test for observed and permuted values was calculated to perform a pairwise comparison among cluster groups. In the following results, we could see that pairs, 1-2, 1-3, and 5-6 were not significantly different but, rest of the pairs were significantly different from each other. From the cluster analysis, we can notice that pairs 1-2 and 1-3 are normal operating conditions of the machine and the statistical test provides sufficient rationale to indicate that there should be no significant difference among these pairs. Pairs 5-6 are the non-normal modes of the machine. Cluster 5 was the displaced shaft, and cluster 6 was imbalanced. In both the cases, the imbalance was noticed, and hence, this test provides sufficient rationale for this case.

A box-plot for clusters and its distance to the centroid was used to identify that all means of clusters are significantly different from each other as shown in Figure 14. In the box plot, we can notice that the cluster centroids are significantly different for each cluster. A two-dimensional principal coordinate analysis (PCoA) was used to study the multivariate dispersion as shown in Figure 15. In PCoA plot, we can observe that cluster 1, cluster 4 and cluster 3 are well separated. While cluster 6 is scattered over cluster 2 and cluster 5. This is a strong indication of the presence of an imbalance in the condition in cluster 4 and cluster 5.

From the results of statistical test and analysis, we have sufficient evidence to conclude that there is a significant difference in the variation between the cluster groups and reject the null hypothesis.

## 5. CONCLUSION

In this paper, vibration and ambient temperature data were collected from an industrial furnace fan for over five months. The data was analyzed using density plots where multi-model distributions were observed. Based on this observation, Gaussian finite mixture model-based clustering was used to cluster the data into distinct groups, and the optimal number of clusters was identified using WSS and elbow technique. In this research, no class labels were available to test the validity of the model. Hence, a hypothesis was developed to identify if, there was any significant difference between the groups (clusters). Different statistical tests such as ANOVA and MANOVA were considered for hypothesis testing. The data were reanalyzed to identify if the data met all the test assumptions. Based on the analyzed results, it was concluded that the identified tests failed to meet the assumptions. Hence, a new technique proposed by Anderson called PERMANOVA along with posthoc pair wise testing was performed to statistically validate the results. Upon performing the PERMANOVA test, it was identified that there was a significant difference among the groups at a p-valueless than the significance level of 0.05.

A test for the analysis of multivariate homogeneity of group dispersions (variances) was also performed. From the results, there was a significant difference between the groups was identified. In pair wise comparison for permuted and observed p-values, all the normal operating conditions had nosignificant difference between them. Likewise, there was no significant difference between the imbalance states of the machine. This research also derived a strong statistical technique to observe if, the maintenance performed on the machines had any significant effect in returning the machine to normal condition. From the following results, we have sufficient evidence to reject the null hypothesis and conclude that there is significant difference among the clusters. Also from the results, we can conclude that by using PERMANOVA testing and pair wise testing we can identify the significance of maintenance in resolving the machine related issues.

## 6. FUTURE SCOPE OF WORK

Using the PERMANOVA test, a statistical difference between the clusters was identified. In machine maintenance, it is important to study what factors affect the formation of each cluster. This research is aimed to be expanded to further to perform factor analysis for determining those critical factors and predicting the time to failure.

## AUTHORS

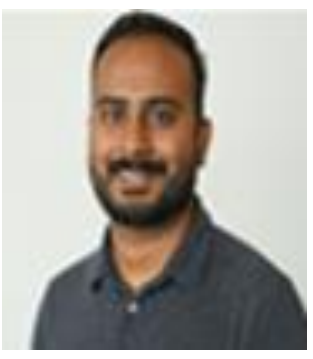

**NagdevAmruthnath** is currently a Ph.D. Student in Industrial Engineering Department, Western Michigan University (WMU). He earned his master’s degree in Industrial engineering from WMU and a Bachelor's degree in Information Science and Engineering from Visvesvaraya Institute of Technology, Karnataka, India. He has four years of experience working in manufacturing industry specializing in the implementation of lean manufacturing and JIT technologies, one year of experience in data analytics, machine learning and AI applications in manufacturing and undergrad teaching experience. The author has journal and conference proceeding publications in production flow analysis, ergonomics, machine learning, and wireless sensor networks. Currently, his research focus is on developing machine learning and AI technologies for manufacturing application. Nagdev continues to serve as a reviewer of scholarly journals which includes Journal of Electrical Engineering and IEEE transactions on Reliability.

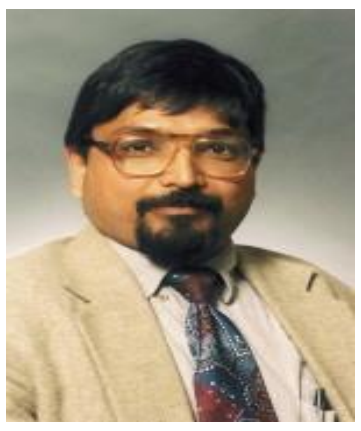

**Dr.Tarun Gupta** is Professor in the Department of Industrial & Entrepreneurial Engineering at Western Michigan University for 29 years. He is a Ph.D. in 1988 from the University of Wisconsin in industrial & systems engineering with a minor in computer science; a 1979 B.Tech in Mechanical Engineering from IITBHU Varanasi, and a 1981 graduate of NITIE, Mumbai India with a master’s Industrial & Systems Engineering. His prime areas of research are manufacturing automation and robotics. He has also continued to research nano-sciences area, machine learning and data analytics applications in advanced manufacturing systems. He has published over 100 papers in his areas of interest. He has also been the recipient of numerous consulting assignments from area industry for specific manufacturing systems challenges. Dr. Gupta has also served as associate editor of International Journal of Robotics and Automation and continues to serve as a reviewer of scholarly journals which includes International Journal of Production Research, International Journal of Nanophotonics, International Journal of Computer & Industrial Engineering, IJCI, & Journal of Operations Management. Dr. Gupta is a member of IEEE, lifetime member of SME and lifetime member of Society of Photonics & Instrumentation Engineers (SPIE), and a past member of IIE, ORSA, ASEE.